\documentclass[prd,showpacs,nofootinbib]{revtex4}
\usepackage{amsmath}
\usepackage{amssymb}
\usepackage{color}
\usepackage{epsf,epsfig}

\def\pr{{Phys. Rev.}~}

\begin{document}
\title{Exclusive processes $e^+e^-\to VP$ in $k_T$ factorization}
\author{Cai-Dian L\"u,  Wei Wang\footnote{Email: wwang@mail.ihep.ac.cn}
 and Yu-Ming Wang\footnote{Email: wangym@mail.ihep.ac.cn}}
\affiliation{ Institute of High Energy Physics, CAS, P.O. Box
918(4), 100049, P.R. China}

\begin{abstract}
The exclusive processes $e^+e^-\to VP$, in the region of which the
final state meson momentum is much larger than the  hadronic scale
$\Lambda_{QCD}$, are studied in the framework of PQCD approach based
on the $k_T$ factorization. Including the transverse momentum
distribution in the light cone wave functions, our results are
consistent with the experimental measurements. According to our
results, many processes have large enough cross sections to be
detected at $\sqrt s=10.58$ GeV. The $s$ dependence of the cross
section has been directly studied and our result indicates that the
$1/s^3$ scaling is more favored than $1/s^4$. We also find that the
gluonic contribution for the processes involving $\eta^{(')}$ is
tiny.
\end{abstract}

\maketitle

\section{Introduction}

In exclusive or inclusive processes with large momentum transfers,
the production rates and many other phenomena, such as the
dimensional rule, the helicity structure, can be successfully
explained by the perturbative QCD analysis \cite{pQCDBL,pQCDER}. The
essential ingredient is the factorization theorem which insures that
a physical amplitude can be represented as a convolution of a hard
scattering kernel and hadron distribution amplitudes. The former can
be calculated using the perturbation theory while the latter,
although non-perturbative in nature, are universal. The light cone
distribution amplitudes which describe the longitudinal momentum
distribution of partons in  hadron, can be determined by the
experiments of various channels. In $e^+e^-\to\gamma^*\to VP$ at
high energies (V denotes a light vector meson and P denotes a light
pseudoscalar meson), the energy of the light meson is much larger
than its mass and the hadronic scale $\Lambda_{QCD}$. One important
feature of this process is that the meson moves nearly on the light
cone. The energetic light meson is composed of two valence quarks
which are both energetic and collinear. The gluon which generates
the quark pair is very hard and this leads to the application of
perturbative QCD into this process. However, the application of the
perturbative QCD approach to this simple process is complicated by
the end point problem. If collinear factorization is applied, the
hard kernel contains the inverse term of momentum fraction which
makes the integration divergent at the end point. This divergence
arises from the overlap of the soft and collinear momentum region
\footnote{This overlap has also been attempted to subtract out in
Ref.~\cite{zerobin} and new factorization theorems in rapidity space
are subsequently achieved.}. A modified perturbative QCD approach
based on $k_T$ factorization, which keeps the intrinsic transverse
momentum of partons in the meson, is proposed and successfully
applied to many processes \cite{PQCDLS,PQCDBdecay}. In this
approach, the Sudakov effect is taken into account and the
applicability of perturbative QCD can be extended down to a few GeV
scale. It is claimed that the perturbative calculations could be
consistent at scale about $Q\sim 20\Lambda_{QCD}$ in this framework.
This approach is also called PQCD approach for simplicity.

The exclusive two-meson  productions in $e^+e^-$ annihilation
provide an opportunity to investigate the behaviors of various meson
form factors. The dependence of the form factor on the energy scale
can shed light on the internal strong interaction information. It
can also give information on  the wave function of the hadrons in
terms of its partonic constituents. In the standard model, the
exclusive production of hadron pairs at $e^+ e^-$ colliders can
proceed through a virtual photon or a $Z^0$ boson. At energies well
below the mass of $Z^0$, the production proceeds predominantly via
the annihilation of $e^+ e^-$ into a virtual photon. Due to the
invariance of charge conjugation in electromagnetic and strong
interactions, the final state should have the same charge
conjugation quantum number with a photon, i.e., these processes can
only produce final states with charge conjugation quantum number
$C=-1$. $e^+e^-\to VP$ can proceed via the following form factor:
\begin{eqnarray}
\langle V(\epsilon,p_1)P(p_2)|j_\mu^{em}|0\rangle
=F_{VP}(s)\epsilon_{\mu\nu\alpha\beta}\epsilon^\nu p_1^\alpha
p_2^\beta,\label{eqs:formfactor}
\end{eqnarray}
where $p_1(p_2)$ is the momentum of the vector (pseudoscalar) meson
and $\epsilon$ is the polarization vector of the vector meson. Here
$j_\mu^{em}$ is defined as $j_\mu^{em}=\bar{q}\gamma_{\mu} q$.
Eq.~(\ref{eqs:formfactor}) indicates that the vector meson is
transversely polarized. On the experimental side, the productions of
$VP$ have been extensively studied:  BES and CLEO-c have reported
the continuum productions \cite{ExpVP1, ExpVP2, ExpVP3}. Recently,
BaBar Collaboration observed the exclusive reaction $e^+e^-\to
\phi\eta$ at $\sqrt s=10.58$GeV and measured the cross section
\cite{phietaBaBar}. Since $k_T$ factorization can give a reliable
prediction in other similar processes, in this paper, we will
perform a study on $e^+e^-\to\gamma^*\to VP$ in this framework and
make a comparison with the data.

\begin{figure}[htb]
\vspace{-0.cm}
\begin{center}
\psfig{file=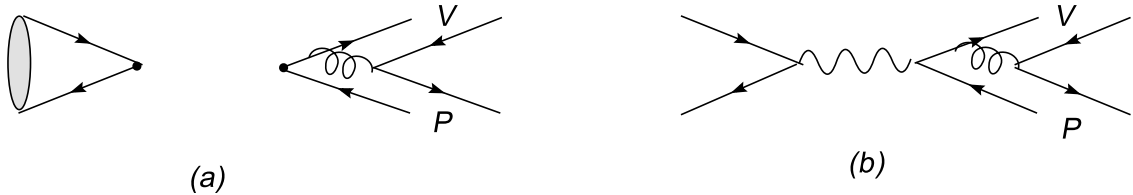,width=15.0cm,angle=0}
\end{center}
\vspace{-0.2cm} \caption{The annihilation diagrams in $B$ decays
and $e^+e^-$ annihilation. In the left diagram, $B$ meson is
annihilated through the four-quark operator. In the right diagram,
electron and positron annihilate into a virtual photon. These two
diagrams have similar topologies.}\label{Bandep}
\end{figure}

Another interesting reason for investigating the $e^+e^-$
annihilation is the similarity with the annihilation corrections in
$B$ decays, shown in  Fig.~\ref{Bandep}. In charmless two body $B$
decays, the annihilation diagrams are power suppressed relative to
the emission contribution. But it is found  that in $B\to \pi K$,
$\pi \pi$ decays, the factorizable annihilation diagrams could be
important due to the chiral enhancement  factor for operator $O_6$
\cite{PQCDBdecay}. This enhancement of factorizable diagrams can
provide large strong phases and give large $CP$ asymmetries, which
indicates the annihilation diagrams are of great importance.
Comparing the two diagrams in Fig. \ref{Bandep}, we can see that the
$e^+e^-$ annihilations have similar topologies with the factorizable
annihilation diagrams in $B$ decays. They may provide an ideal
laboratory to isolate the power correction effect and to find out
whether contributions of annihilations from end-point are important
or not for meson productions \cite{VPandB}.

The remainder of this paper is organized as follows. In
Section~\ref{fram}, we present the expression for the cross sections
for $e^+e^-\to VP$ in the $k_T$ factorization: the first part of
this section is devoted to the discussion on the decay constants and
the distribution amplitudes of mesons, and the second part
contributes to a brief introduction to PQCD approach and
factorization formulae for form factors. The numerical results and
discussions are presented in Section \ref{num}. The last section is
our summary.

\section{Calculation in $k_T$ factorization}\label{fram}

\subsection{Decay constants and Wave functions}

The decay constants for a pseudoscalar meson and a vector meson
are defined by:
\begin{eqnarray}
\langle P(P)|\bar q_2\gamma_\mu\gamma_5
q_1|0\rangle=-if_PP_\mu,\;\; \langle V(P,\epsilon)|\bar
q_2\gamma_\mu q_1|0\rangle=f_Vm_V\epsilon_\mu,\;\;\langle
V(P,\epsilon)|\bar q_2\sigma_{\mu\nu}
q_1|0\rangle=-if_V^T(\epsilon_\mu P_\nu-\epsilon_\nu P_\mu),
\end{eqnarray}
The pseudoscalar decay constants taken from the Particle Data
Group~\cite{PDG} are shown in Table~\ref{decayconstant}. The charged
vector meson longitudinal decay constants are extracted from the
data on $\tau^- \to (\rho^-,K^{*-}) \nu_\tau$, while the neutral
vector meson longitudinal decay constants are determined from the
data on the electromagnetic annihilation processes $V^0\to
e^+e^-$~\cite{PDG}. The transverse decay constants are taken from
the QCD sum rules~\cite{LCSRBZ,adoptedvectorwf}, which are also
collected in Table~\ref{decayconstant}.

\begin{table} \caption{Input values of the decay constants  of the pseudoscalar and vector
mesons (in MeV)~\cite{PDG,LCSRBZ,adoptedvectorwf}}
\begin{tabular}{cccccccccc}
\hline\hline
 $f_\pi $ & $f_K $ & $f_\rho $ & $ f_\rho^T $ & $ f_\omega $ & $ f_\omega^T $
 & $ f_{K^*} $ & $ f_{K^*}^T $ & $f_\phi $ & $
f_\phi^T $  \\
131 & 160& $ 209 $& $ 165\pm 9 $&
 $ 195 $&
 $ 145\pm 10$&
$ 217 $&
 $185\pm 10$&
 $ 231 $&
 $ 200\pm 10$\\
\hline \hline
\end{tabular}\label{decayconstant}
 \end{table}

The light-cone distribution amplitudes are defined by the matrix
elements of the non-local operators at the light-like separations
$z_\mu$ with $z^2=0$, and sandwiched between the vacuum and the
meson state. The two-particle light-cone distribution amplitudes of
an outgoing  pseudoscalar meson $P$, up to twist-3 accuracy, are
defined by \cite{PseudoscalarWV}:
\begin{eqnarray}
\langle P(P)|{\bar q}_{2\beta}(z)q_{1\alpha}(0)|0\rangle &=&
-\frac{i}{\sqrt{6}}\int_0^1 dx e^{ixP\cdot z} \left[\gamma_5\not\!
P\phi^A(x) +m_0\gamma_5\phi^P(x) -m_0\sigma^{\mu\nu}\gamma_5
P_{\mu}z_{\nu} \frac{\phi^{\sigma}(x)}{6}\right]_{\alpha\beta}  \nonumber\\
&=&-\frac{i}{\sqrt{6}} \int_0^1dx e^{ixP\cdot z}\left[\gamma_5\not
\! P\phi^A(x) + \gamma_5m_0\phi^P(x) +m_0\gamma_5(\not \! n\not
\!v-1) \phi^T(x)\right]_{\alpha\beta}\;,  \label{fpd}
\end{eqnarray}
where $n,v$ are two light-cone vectors. The pseudoscalar meson is
moving on the direction of $n$, with $v$ the opposite direction.
$m_0= \frac{M_P^2}{m_{q_1}+m_{q_2}}$ is the chiral enhancement
parameter. $x$ is the momentum fraction carried by the positive
quark $q_2$. We have performed the integration by parts for the
third term and $\phi^T(x)=\frac{1}{6}{d \over dx}\phi^\sigma(x)$.
The explicit form of distribution amplitudes for pseudoscalar mesons
have been studied in QCD sum rule approach and other methods
\cite{PseudoscalarUpdate,NewPWV}. In principle, they are
factorization scale dependent. Here we use  the following form for
leading twist distribution amplitudes
\begin{eqnarray}
\phi_{\pi}^A(x) &=& \frac{3f_{\pi}}{\sqrt{6}} x(1-x)[ 1 +a_2^\pi C_2^{3/2}(t)], \\
\phi_{K}^A(x) &=&
\frac{6f_{K}}{2\sqrt{6}}x(1-x)[1+a_1^KC_1^{3/2}(t)+a_2^KC_2^{3/2}(t)],
\end{eqnarray}
where $t=2x-1$ and Gegenbauer polynomials defined as:
\begin{equation}
C^{3/2}_{1}(t)=3t, C_2^{3/2} (t)=\frac{3}{2} (5t^2-1).
\end{equation}
The Gegenbauer moments at $\mu=1$ GeV are determined as
\begin{eqnarray}
a_2^{\pi,K}=0.25\pm0.15, \;\;\; a_1^K=0.06\pm 0.03.
\end{eqnarray}
Since the momentum transfer at $\sqrt s=10.58$ GeV is large enough,
the use of asymptotic forms for twist-3 distribution amplitudes  is
acceptable. Besides, we also use these forms at $\sqrt s=3.67$ GeV
for simplicity. The asymptotic forms of twist-3 distribution
amplitudes are given as:
\begin{eqnarray}
\phi_{\pi(K)}^P(x) &=& \frac{f_{\pi(K)}}{2\sqrt{6}},\;\;\;
\phi_{\pi(K)}^T(x) = \frac{f_{\pi(K)}}{2\sqrt{6}}(1-2x).
\end{eqnarray}

As for the mixing of $\eta$ and $\eta^\prime$, we use the quark
flavor basis proposed by Feldmann and Kroll \cite{FKmixing}, i.e.
these two mesons are made of $\bar nn=(\bar uu+\bar dd)/\sqrt 2$
and $\bar ss$:
\begin{equation}
   \left( \begin{array}{c}
    |\eta\rangle \\ |\eta^\prime\rangle
   \end{array} \right)
   = U(\theta)
   \left( \begin{array}{c}
    |\eta_{n}\rangle \\ |\eta_s\rangle
   \end{array} \right),
\end{equation}
with the mixing matrix,
\begin{equation}
U(\theta)=\left( \begin{array}{cc}
    \cos\theta & ~-\sin\theta \\
    \sin\theta & \phantom{~-}\cos\theta
   \end{array} \right)\;,
\end{equation}
where the mixing angle $\theta=39.3^\circ\pm1.0^\circ$. In
principle, this mixing mechanism is equivalent to the singlet and
octet formalism, which is shown in \cite{GluonicCKL}. But the
advantage is transparent, since only two decay constants are needed:
\begin{eqnarray}
   \langle 0|\bar n\gamma^\mu\gamma_5 n|\eta_n(P)\rangle
   &=& \frac{i}{\sqrt2}\,f_n\,P^\mu \;,\nonumber \\
   \langle 0|\bar s\gamma^\mu\gamma_5 s|\eta_s(P)\rangle
   &=& i f_s\,P^\mu \;.\label{deffq}
\end{eqnarray}

We assume that the wave function of $\bar{n}n$ and $\bar ss$ is the
same as the pion's wave function, except for the different decay
constants and the chiral scale parameters:
\begin{eqnarray}
 f_n=(1.07\pm0.02)f_{\pi}, \ \ \ f_s=(1.34\pm0.06)f_\pi.
 \end{eqnarray}
The chiral enhancement factors are chosen as
\begin{eqnarray}
  m_0^{\bar {n} n}=\frac{1}{2m_n}[m^2_\eta\cos^2\theta+
  m_{\eta'}^2\sin^2\theta-\frac{\sqrt 2f_s}{f_n}(m_{\eta'}^2-m_\eta^2)\cos\theta\sin\theta],\\
  m_0^{\bar{s} s}=\frac{1}{2m_s}[m^2_{\eta'}\cos^2\theta+
  m_{\eta}^2\sin^2\theta-\frac{f_n}{\sqrt
  2f_s}(m_{\eta'}^2-m_\eta^2)\cos\theta\sin\theta],
\end{eqnarray}
with $m_n=5.6$ MeV and $m_s=137$ MeV at $\mu=1$ GeV \cite{NewPWV}.

In this work, we also investigate the gluonic contribution  for
iso-singlet pseudoscalar meson $\eta$ and $\eta'$. This contribution
has been attempted in \cite{GluonicCKL} with negligible effect in
$B\to\eta$ form factor and a few percents to $B\to \eta'$. The
leading-twist gluonic distribution amplitudes of the $\eta_n$ and
$\eta_s$ mesons are defined as \cite{leadingtwistofeta}:
\begin{eqnarray}
\langle\eta_n(P)|A_{[\mu}^a(z)A_{\nu]}^b(0)|0\rangle
&=&\frac{\sqrt{2}f_n}{\sqrt{3}}\frac{C_F}{4\sqrt{3}}
\frac{\delta^{ab}}{N_c^2-1}\epsilon_{\mu\nu\rho\sigma}\,
\frac{n_-^\rho P^\sigma\,}{n_-\cdot P} \int_0^1dx
e^{ixP\cdot z}\frac{\phi_{n}^{G}(x)}{x(1-x)} \;,\nonumber\\
\langle\eta_s(P)|A_{[\mu}^a(z)A_{\nu]}^b(0)|0\rangle
&=&\frac{f_s}{\sqrt{3}}\frac{C_F}{4\sqrt{3}}
\frac{\delta^{ab}}{N_c^2-1}\epsilon_{\mu\nu\rho\sigma}\,
\frac{n_-^\rho P^\sigma\,}{n_-\cdot P} \int_0^1dx e^{ixP\cdot
z}\frac{\phi_{s}^{G}(x)}{x(1-x)} \;, \label{qgn}
\end{eqnarray}
where $A_{[\mu}^a(z)A_{\nu]}^b(w)\equiv
[A_{\mu}^a(z)A_{\nu}^b(w)-A_{\nu}^a(z)A_{\mu}^b(w)]/2$ and the
function \cite{Gegenbauerofeta},
\begin{eqnarray}
\phi_{n(s)}^{G}(x)=x^2(1-x)^2
B_2^{n(s)}C_1^{5/2}(2x-1)\;,\;\;\;\;C_1^{5/2}(t)=5t\;.
\label{phiG}
\end{eqnarray}
The gluon labeled by the subscript $\mu$ carries the momentum
fractions $x$ based on the above definition. The two Gegenbauer
coefficients $B_2^n$ and $B_2^s$ could not be the same in principle.
However, it is acceptable to assume $B_2^n=B_2^s\equiv B_2$, since
there are large uncertainties in their values. Here the range of
$B_2$ has been extracted as $ B_2=4.6\pm2.5$
\cite{leadingtwistofeta}.

Following the similar procedures as for the pseudoscalar mesons,
we can derive the vector meson distribution amplitudes for the
transverse polarization up to twist-3 \cite{rho}:
\begin{eqnarray}
 \langle V(P,\epsilon_T)|\bar q_{2\beta}(z)
q_{1\alpha}(0)|0\rangle &=&\frac{1}{\sqrt{6}}\int_0^1 dx
e^{ixP\cdot z} \left[ M_V\not\! \epsilon_T\phi_V^v(x)+
\not\!\epsilon_T\not\! P\phi_V^T(x)\right.
\nonumber\\
& & \left.+M_V
i\epsilon_{\mu\nu\rho\sigma}\gamma_5\gamma^\mu\epsilon_T^\nu n^\rho
v^\sigma \phi_V^p(x)\right ]_{\alpha\beta}, \; \label{spf}
\end{eqnarray}
where we have adopted the convention $\epsilon^{0123}=1$ for the
Levi-Civita tensor $\epsilon^{\mu\nu\alpha\beta}$.

The twist-2 distribution amplitudes for transversely polarized
vector can be expanded as:
\begin{eqnarray}
\phi^T_\rho (x)&=&\frac{3f^T_\rho}{\sqrt{6}} x (1-x)\left[1+
a_{2\rho}^{\perp}C_2^{3/2} (t) \right]\;,\label{phirho}\\
\phi^T_\omega(x)&=&\frac{3f^T_\omega}{\sqrt{6}} x (1-x)\left[1+
a_{2\omega}^{\perp}C_2^{3/2} (t)\right]\;,\\
\phi^T_{K^*}(x)&=&\frac{3f^T_{K^*}}{\sqrt{6}} x
(1-x)\left[1+a_{1K^*}^{\perp}C^{3/2}_1(t)+
a_{2K^*}^{\perp}C_2^{3/2} (t)\right]\;,\\
\phi^T_{\phi}(x)&=&\frac{3f^T_{\phi}}{\sqrt{6}} x (1-x)\left[1+
a_{2\phi}^{\perp}C_2^{3/2} (t)\right]\;.\label{phiphi}
\end{eqnarray}
The Gegenbauer moments have been studied extensively in the
literature \cite{rho,previousvectorwf}, here we adopt the very
recent updated form \cite{adoptedvectorwf,LCSRBZ}:
\begin{eqnarray}
a_{1K^*}^\perp=0.04\pm0.03,\;\;
a_{2\rho}^{\perp}=a_{2\omega}^{\perp}=0.15\pm0.07,\;\;
a_{2K^*}^{\perp}=0.11\pm0.09,\;\;a_{2\phi}^{\perp}=0.06^{+0.09}_{-0.07}.
\end{eqnarray}
As for the twist-3 distribution amplitudes $\phi^p_V$ and
$\phi^v_V$, there is no recent update associate with those updates
for twist-2 distribution amplitudes \cite{adoptedvectorwf,LCSRBZ},
we also use the asymptotic form:
\begin{eqnarray}
\phi_V^v(x)&=&\frac{3f_V}{8\sqrt6}[1+(2x-1)^2],\;\;\; \ \ \
 \phi_V^p(x)=\frac{3f_V}{4\sqrt6}(1-2x).
 \label{twist3vector}
\end{eqnarray}

The above discussions  concentrated on the longitudinal momentum
distribution and we intend to include the transverse momentum
distribution functions of the pseudoscalar and vector mesons. But at
present, the intrinsic transverse momentum dependence of wave
function is still unknown from the first principle of QCD. As an
illustration, we use a simple model in which the dependence of the
wave function on the longitudinal and transverse momentum can be
factorized into two parts \cite{transverseDA}:
\begin{eqnarray}
\psi(x,{\bf{k}}_T)=\phi(x)\times \Sigma({\bf{k}}_T),
\end{eqnarray}
where $\phi(x)$ is the longitudinal momentum distribution
amplitude which has been discussed above and $\Sigma({\bf{k}}_T)$
describes the transverse momentum distribution.
$\Sigma({\bf{k}}_T)$ satisfy the normalization conditions:
\begin{eqnarray}
\int d^2{\bf{k}}_T \Sigma({\bf{k}}_T)=1.
\end{eqnarray}
In the following, we will use a Gaussian distribution:
\begin{eqnarray}
\Sigma({\bf{k}}_T)=\frac{\beta^2}{\pi}\mbox{exp}(-\beta^2{\bf{k}}_T^2),
\end{eqnarray}
where the parameter $\beta$ characterizes the shape of the
transverse momentum distribution. The numerical value for $\beta$
can be fixed by the condition that the root mean square transverse
momentum $\langle {\bf{k}}_T^2\rangle^{1/2}$ should be at the order
of $\Lambda_{QCD}$. Their relation can be derived from:
\begin{eqnarray}
\langle {\bf{k}}_T^2\rangle=\frac{\int^1_0dx\int d^2{\bf{k}}_T
{\bf{k}}_T^2|\psi(x,{\bf{k}}_T)|^2}{\int^1_0dx\int d^2{\bf{k}}_T
|\psi(x,{\bf{k}}_T)|^2}=\frac{1}{2\beta^2}.
\end{eqnarray}
If we choose the root mean square transverse momentum $\langle
{\bf{k}}_T^2\rangle^{1/2}=0.35$ GeV, then $\beta^2= 4
\mbox{GeV}^{-2}$. In PQCD approach, the integration will be
transformed to the ${\bf{b}}$ space (coordinate space) and it is
convenient to use the Fourier transformation of
$\Sigma({\bf{k}}_T)$:
\begin{eqnarray}
\Sigma({\bf{b}})=\int d^2{\bf{k}}_T e^{-i{\bf{k}}_T\cdot
{\bf{b}}}\Sigma({\bf{k}}_T)=\mbox{exp}(-\frac{b^2}{4\beta^2}).\label{transverse}
\end{eqnarray}
It can be observed that in the limit $\beta\to\infty$,
$\Sigma({\bf{b}})$ can be simply replaced by 1.

\subsection{Form factor and cross section in $k_T$ factorization}

In the center of mass frame, we define $q_1$, $q_2$, $p_1$ and $p_2$
to be the four-momenta of $e^{+}$, $e^{-}$ in initial states, vector
(V) and pseudoscalar meson (P) in final states, and define
$k_{1(2)}$ and $x_{1(2)}$ to be the momenta and momentum fractions
of the positive quarks inside V  and P respectively. The center mass
energy of this process is denoted as $Q=\sqrt s$. Using the
definition of the form factor in Eq.(1), we can obtain the cross
section as
\begin{eqnarray}
\sigma(e^+e^-\to VP)=\frac{\pi\alpha_{em}^2}{6}|F_{VP}|^2
\Phi^{3/2}(s),
\end{eqnarray}
with
\begin{eqnarray}
\Phi(s)=\bigg[1-\frac{(m_V+m_P)^2}{s}\bigg]
\bigg[1-\frac{(m_V-m_P)^2}{s}\bigg].
\end{eqnarray}

\begin{figure}[htb]
\vspace{-0.cm}
\begin{center}
\psfig{file=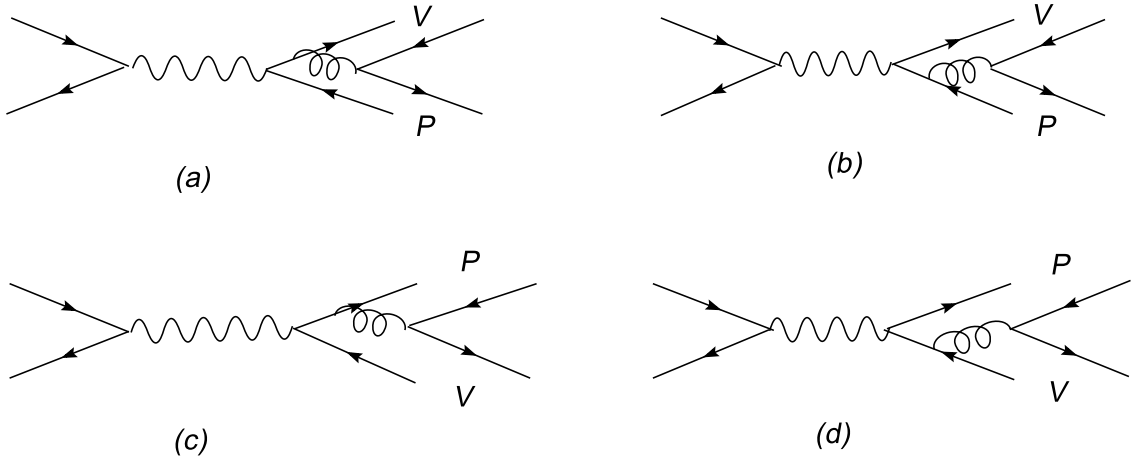,width=15.0cm,angle=0}
\end{center}
\vspace{-0.2cm} \caption{Dominant contribution of $e^+e^-\to
VP$}\label{quark}
\end{figure}

\begin{figure}[htb]
\vspace{-0.cm}
\begin{center}
\psfig{file=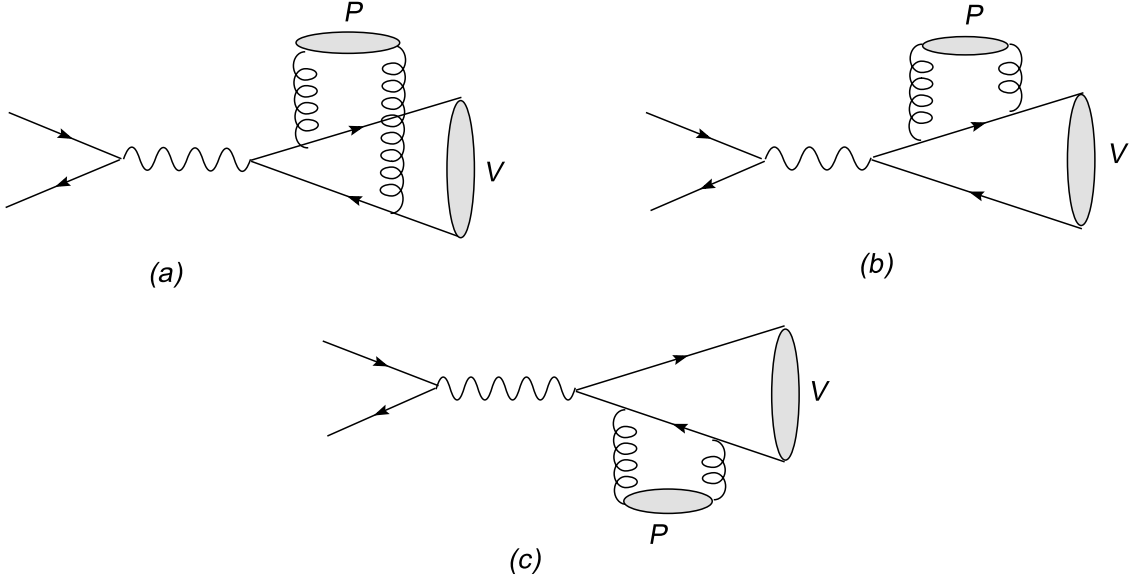,width=15.0cm,angle=0}
\end{center}
\vspace{-0.2cm} \caption{Gluonic contributions}\label{gluonic}
\end{figure}

\begin{figure}[htb]
\vspace{0.3cm}
\begin{center}
\psfig{file=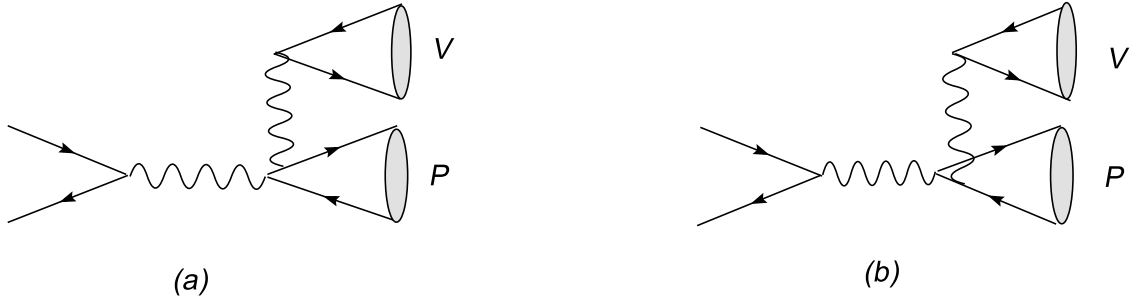,width=15.0cm,angle=0}
\end{center}
\vspace{-0.2cm} \caption{Enhanced diagrams for a neutral vector
meson production}\label{VMD}
\end{figure}

\begin{figure}[htb]
\vspace{0.3cm}
\begin{center}
\psfig{file=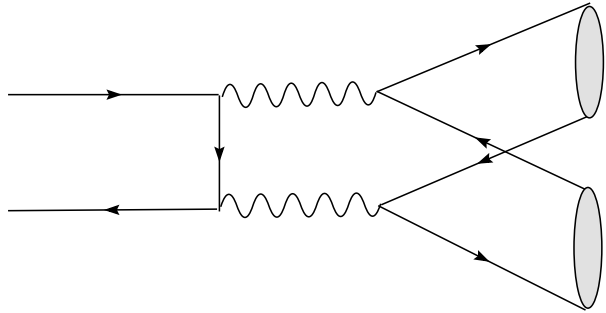,width=8.0cm,angle=0}
\end{center}
\vspace{-0.2cm} \caption{Two photon non-fragmentation diagram.
This contribution is suppressed by $\alpha_{em}$ and can be
neglected.}\label{nonf}
\end{figure}

There  are four different types of diagrams contributing to the
production of vector and pseudoscalar meson in $e^{+}e^{-}$
annihilations, to the leading order of the strong and
electromagnetic coupling constants. The first type of diagrams
contributing to this process are displayed in Fig.~\ref{quark}.
These diagrams give the dominant contribution. The diagrams in
Fig.~\ref{gluonic} contribute to the processes involving $\eta$
and $\eta'$, while the diagrams in Fig.~\ref{VMD} only contribute
to the processes involving a neutral vector meson, such as
$\rho^0$, $\omega$ and $\phi$. Although these diagrams are
suppressed by $\alpha_{em}$, they can be enhanced by
$s/\Lambda_{QCD}^2$. This mechanism is similar with the
enhancement in penguin-dominated $B$ decays \cite{enhancement} and
the so-called fragmentation mechanism in $e^+e^-\to VV$ processes
\cite{fragementation}. It is also interesting to explore this
effect in $e^+e^-\to VP$. For $e^+e^-\to K^*K$ and
$e^+e^-\to\rho^+\pi^-$, the two photon non-fragmentation diagrams
can give their contributions as in Fig.~\ref{nonf}. But these
diagrams suffering the suppression from electromagnetic coupling
constant $\alpha_{em}$ which can be neglected safely.

We begin with a brief review of the PQCD approach. The basic idea of
PQCD approach is that it takes into account the transverse momentum
of valence quarks which results in the Sudakov factor. The form
factor, taking the first diagram in Fig.~\ref{quark}  as an example,
can be expressed as the convolution of the wave functions $\psi_V$,
$\psi_P$ and the hard scattering kernel $T_H$ by both the
longitudinal and the transverse momenta:
\begin{eqnarray}
F_{a}(VP)=\int^1_0dx_1dx_2\int
d^2{\bf{k}}_{T_1}d^2{\bf{k}}_{T_2}\psi_V(x_1,{\bf{k}}_{T_1},p_1,\mu)
T_H(x_1,x_2,Q,{\bf{k}}_{T_1},{\bf{k}}_{T_2},\mu)
\psi_P(x_2,{\bf{k}}_{T_2},p_2,\mu).
\end{eqnarray}
Through the Fourier transformation, the above equation can be
expressed as:
\begin{eqnarray}
F_{a}(s)=\int^1_0dx_1dx_2\int
\frac{d^2{\bf{b}}_{1}}{(2\pi)^2}\frac{d^2{\bf{b}}_{2}}{(2\pi)^2}{\cal
P}_V(x_1,{\bf{b}}_{1},p_1,\mu)
T_H(x_1,x_2,Q,{\bf{b}}_{1},{\bf{b}}_{2},\mu){\cal
P}_P(x_2,{\bf{b}}_{2},p_2,\mu).
\end{eqnarray}
Here ${\cal P}_{i}(x_j,{\bf{b}}_{j},p_j,\mu)$ are the Fourier
transformation of $\psi_{i}(x_j,{\bf{k}}_{T_j},p_j,\mu)$, where the
subscript $i$ denotes $V$ or $P$, and $j$ indicates $1$ or $2$.

In the above expression, the double logarithms, arising from the
overlap of the soft and collinear divergence, have been resummed to
result in the Sudakov factor \cite{Sudakov}
\begin{equation}
{{\cal
P}_i}(x_j,{\bf{b}}_j,p_j,\mu)=\exp\left[-s(x_j,b_j,Q)-s(1-x_j,b_j,Q)\right]
{\bar{\cal P}_i}(x_j,{\bf{b}}_j,\mu)\; . \label{sp}
\end{equation}
The exponent $s(\xi,b_j,Q)$, $\xi=x_j$ or $1-x_j$, is expressed as
\begin{equation}
s(\xi,b_j,Q)=\int_{1/b_j}^{\xi Q/\sqrt{2}}\frac{d p}{p}
\left[\ln\left(\frac{\xi Q}
{\sqrt{2}p}\right)A(\alpha_s(p))+B(\alpha_s(p))\right]\;,
\label{fsl}
\end{equation}
where the anomalous dimensions $A$ and $B$ to one loop are given by
\begin{eqnarray}
A&=&{ C}_F\frac{\alpha_s}{\pi}, \,\,\,
B=\frac{2}{3}\frac{\alpha_s}{\pi}\ln\left(\frac{e^{2\gamma_E-1}}
{2}\right)\;,
\end{eqnarray}
with $C_F={N_c^2-1 \over 2 N_c}$ and $\gamma_E$ being the Euler
constant. The one-loop running coupling constant,
\begin{equation}
\frac{\alpha_s(\mu)}{\pi}=\frac{4}{\beta_0\ln(\mu^2/\Lambda_{QCD}^2)}\;,
\label{ral}
\end{equation}
with the coefficients
\begin{eqnarray}
& &\beta_{0}=\frac{33-2n_{f}}{3}, \label{12}
\end{eqnarray}
where $n_f$ is the number of the active quark number. We require the
relation of the involved scales $\xi Q/\sqrt{2}> 1/b_j > \Lambda$ as
indicated by the bounds of the variable $p$ in Eq.~(\ref{fsl}). The
QCD dynamics below $1/b_j$ scale is regarded as being
nonperturbative which can be absorbed into the initial condition
${\bar{\cal P}_i}(x_j,{\bf{b}}_j,\mu)$.

The form factor, as a physical observable, is independent of
renormalization scale $\mu$, but the functions ${\bar{\cal P}}$ and
$T_H$ still contain single logarithms from ultraviolet divergences,
which can be summed using the renormalization group equation method.
This renormalization group analysis applied to $T_H$ gives
\begin{eqnarray}
{T_H}(x_j,{\bf{b}}_j,Q,\mu)
&=&\exp\left[-4\,\int_{\mu}^{t}\frac{d\bar{\mu}}{\bar{\mu}}
\gamma_q(\alpha_s(\bar{\mu}))\right]\times
{T_H}(x_j,{\bf{b}}_j,Q,t)\;, \label{13}
\end{eqnarray}
where $\gamma_q=-\alpha_s/\pi$ is the quark anomalous dimension in
axial gauge and $t$ is the largest mass scale involved in the hard
scattering,
\begin{equation}
{t=\max(\sqrt{x_{2}}Q,1/b_1,b_2)}. \label{hardscale1}
\end{equation}
{The scale $\sqrt{x_2}Q$ is associated with the longitudinal
momentum of the quark propagator and $1/b_j$ with the transverse
momentum.} The large-$b_j$ behavior of $\cal P$ is summarized as
\begin{eqnarray}
{\cal P}_i
(x_j,{\bf{b}}_j,p_j,\mu)&=&\exp\left[-s(x_j,b_j,Q)-s(1-x_j,b_j,Q)
-2\int_{1/b_j}^{\mu} \frac{d\bar{\mu}}{\bar{\mu}}\gamma
_q(\alpha_s(\bar{\mu}))\right] \times {\bar{\cal
P}}_i(x,{\bf{b}}_j,1/b_j),\; \label{pb}
\end{eqnarray}
where ${\bar{\cal P}}_i(x_j,{\bf{b}}_j,1/b_j)$ is the wave function
discussed above: \begin{eqnarray} {\bar{\cal
P}}_i(x_j,{\bf{b}}_j,1/b_j)=\phi(x_j,1/b_j)\times
\Sigma({\bf{b}}_j).
\end{eqnarray}

The threshold resummation \cite{threshold,thresholdB} can also play
an important role in $e^+e^-\to VP$ processes. The lowest-order
diagrams Fig.~\ref{quark}(a) and (d) give an amplitude proportional
to $1/(x_2^2  (1-x_1))$ and $1/(x_2^2 x_1)$ respectively. In the
threshold region with $x_2\to 0$  {(to be precise, $x_2\sim
O(\Lambda_{QCD}^2/s)$)}, additional collinear divergences are
associated with the internal quark. The QCD loop correction to the
electromagnetic vertex can produce the double logarithm
$\alpha_s\ln^2 x_2$ and  resummation of this type of double
logarithms lead to the Sudakov factor $S_t(x_2)$. Similarly,
resummation of $\alpha_s\ln^2 x_1$ due to loop corrections in the
other diagrams lead to the Sudakov factor $S_t(x_1)$. The Sudakov
factor from threshold resummation is universal, independent of
flavors of internal quarks, twists,  and the specific processes. To
simplify the analysis, the following parametrization has been used
\cite{thresholdB}:
\begin{eqnarray}
S_t(x)=\frac{2^{1+2c}\Gamma(3/2+c)}{\sqrt{\pi}\Gamma(1+c)}
[x(1-x)]^c\;, \label{str}
\end{eqnarray}
with the parameter $c=0.3$. This parametrization, symmetric under
the interchange of $x$ and $1-x$, is convenient for evaluation of
the amplitudes. It is obvious that the threshold resummation
modifies the end-point behavior of the meson distribution
amplitudes, rendering them vanish faster at $x\to 0$.

Combing all the above ingredients, we obtain the factorization
formula for the contribution from Fig.~\ref{quark}(a):
\begin{eqnarray}
F_a(VP)&=&16\pi C_F Q r_1\int_0^1 dx_1dx_2\int_0^\infty b_1db_1b_2db_2\nonumber\\
&&\times \phi_P^A(x_2,b_2)[\phi_V^p(x_1,b_1)-\phi_V^v(x_1,b_1)]
E(t_a)h(1-x_1,x_2,b_1,b_2),
\end{eqnarray}
where $h$ and $E$ are defined by \cite{PQCDBdecay}
\begin{eqnarray}
h(x_1,x_2,b_1,b_2)&=&(\frac{i\pi}{2})^2
S_t(x_2)\Big[\theta(b_1-b_2)H_0^{(1)}(\sqrt{x_2}Qb_1)J_0(\sqrt
{x_2}Qb_2)\nonumber\\
&&\;\;+\theta(b_2-b_1)H_0^{(1)}(\sqrt {x_2}Qb_2)J_0(\sqrt
{x_2}Qb_1)\Big]H_0^{(1)}(\sqrt{x_1x_2}Qb_1),\\
E(t_a)&=&\alpha_s(t_a)\mbox{exp}[-S_1(t_a)-S_2(t_a)],
\end{eqnarray}
where $J_0$ and $H_0^{(1)}$ are the Bessel   functions,
respectively, $t_a=\max(\sqrt{x_{2}}Q,1/b_1,1/b_2)$ and $r_1=M_V/Q$.

Similarly, for the other diagrams, the amplitudes are:
\begin{eqnarray}
F_b(VP)&=&-16\pi C_F Q\int_0^1 dx_1dx_2\int_0^\infty b_1db_1b_2db_2E(t_b)h(x_2,1-x_1,b_2,b_1)\nonumber\\
&&\times
\Big\{r_1(x_1-1)[\phi_1^p(x_1,b_1)+\phi_1^v(x_1,b_1)]\phi_2^A(x_2,b_2)+2r_2\phi_1^T(x_1,b_1)\phi_2^P(x_2,b_2)\Big\}
,\\
F_c(VP)&=&-16\pi C_F Q\int_0^1 dx_1dx_2\int_0^\infty b_1db_1b_2db_2E(t_c)h(1-x_2,x_1,b_2,b_1)\nonumber\\
&&\times
[r_1x_1(\phi_1^p(x_1,b_1)-\phi_1^v(x_1,b_1))\phi_2^A(x_2,b_2)+2r_2\phi_1^T(x_1,b_1)\phi_2^P(x_2,b_2)]
,\\
F_d(VP)&=&-16\pi C_F Q r_1\int_0^1 dx_1dx_2\int_0^\infty b_1db_1b_2db_2E(t_d)h(x_1,1-x_2,b_1,b_2)\nonumber\\
&&\times (\phi_1^p(x_1)+\phi_1^v(x_1))\phi_2^A(x_2,b_2),
\end{eqnarray}
with $r_2=m_0/Q$. The factorization scales $t_i$ are chosen as
\begin{eqnarray}
t_b=\max(\sqrt{1-x_{1}}Q,1/b_1,1/b_2),\;\;\;
t_c=\max(\sqrt{x_{1}}Q,1/b_1,1/b_2),\;\;
t_d=\max(\sqrt{1-x_{2}}Q,1/b_1,1/b_2).\label{hardscale2}
\end{eqnarray}

If the final state meson is not $K^*$ or $K$, the distribution
amplitudes are completely symmetric or antisymmetric under the
interchange of $x_j$ and $1-x_j$. Then one can easily obtain:
\begin{eqnarray}
F_a(VP)=F_d(VP),\;\;\; F_b(VP)=F_c(VP).\label{eq:relation}
\end{eqnarray}

For the flavor-singlet pseudoscalar meson $\eta$ and $\eta'$, there
are additional contributions from the two-gluon diagrams as
displayed in Fig.~\ref{gluonic}, even if they may be suppressed by
the gluonic distribution amplitudes. However, it is still worthwhile
to investigate the numerical contribution in order to make our
calculations as complete as possible. The computations of these
diagrams are much similar with that showed in Fig.~\ref{quark}. The
explicit calculations show that the Fig.~\ref{gluonic} (a) does not
contribute to the transition amplitude, due to the antisymmetry of
the two gluons. The amplitudes of the other two diagrams are given
as
\begin{eqnarray}
F_e(V\eta_s)&=&-8\pi Q r_1\frac{f_{s}C_F^2\sqrt {2N_c}}{3(N_c^2-1)}
\int_0^1 dx_1dx_2\int_0^\infty
b_1db_1b_2db_2E(t_e)h(x_2,x_1,b_1,b_2)\nonumber\\
&&\times
[(x_2+1)\phi_V^v(x_1,b_1)-(x_2-1)\phi_V^p(x_1,b_1)]\frac{\phi_s^G(x_2,b_2)}{x_2(1-x_2)}
,\label{Fe}\\
\ \ F_f(V\eta_s)&=&8\pi Q r_1\frac{f_{s}C_F^2\sqrt
{2N_c}}{3(N_c^2-1)}
\int_0^1 dx_1dx_2\int_0^\infty b_1db_1b_2db_2E(t_f)h(1-x_2,1-x_1,b_1,b_2)\nonumber\\
&&\times
(x_1-1)[(x_2-2)\phi_V^v(x_1,b_1)-x_2\phi_V^p(x_1,b_1)]\frac{\phi_s^G(x_2,b_2)}{x_2(1-x_2)},
\label{Ff}
\end{eqnarray}
for $e^+e^-\to V\eta_s$ process with
\begin{eqnarray}
t_e=\max(\sqrt{x_{1}}Q,1/b_1,1/b_2),\;\;\;
t_f=\max(\sqrt{1-x_{1}}Q,1/b_1,1/b_2).\label{hardscale3}\end{eqnarray}
It should be pointed out that the factor ``2" from the exchange of
two identical gluons in the final states has been added in the
above equations. The amplitude for $e^+e^-\to V\eta_n$ can be
easily obtained by replacing the corresponding decay constant with
an additional factor $\sqrt 2$ from Eq. (\ref{Fe},\ref{Ff}).

Furthermore, there are also contributions from the transition of
photon radiated from one valence quark in pseudoscalar meson into
vector meson directly, which have been presented in
Fig.~{\ref{VMD}}. Although these diagrams maybe suppressed by
coupling constant of electromagnetic interactions, they are also
enhanced by the almost on-shell photon propagator compared with the
first type diagrams, especially for the processes with a very large
center mass energy. These diagrams can also be calculated according
to $k_T$ factorization, however, we will simply adopt collinear
factorization due to disappearance of infrared divergence for these
two diagrams. These two amplitudes are equal after integrating the
momentum fractions carried by the valence quark of the meson. Hence,
we obtain the amplitudes corresponding to them as follows:
\begin{eqnarray}
F_g(VP)&=&F_h(VP)=\frac{12\pi \alpha_{em}f_Vf_P}{M_V s}(1+a_2^P).
\end{eqnarray}

The form factors for the explicit channels can be easily obtained
from the combinations of the eight amplitudes $F_{a-h}$. To be more
specific, we can write them as
\begin{eqnarray}
F_{\rho^+\pi^-}&=&\frac{1}{3}[F_a(\rho\pi)+F_b(\rho\pi)],\label{rhopi1}\\
F_{\rho^0\pi^0}&=&\frac{1}{3}[F_a(\rho\pi)+F_b(\rho\pi)]+\frac{1}{6}[F_g(\rho\pi)+F_h(\rho\pi)],\label{rhopi2}\\
F_{\omega\pi^0}&=&[F_a(\omega\pi)+F_b(\omega\pi)]+\frac{1}{18}[F_g(\rho\pi)+F_h(\rho\pi)],\\
F_{\phi\pi^0}&=&-\frac{\sqrt 2}{18}[F_g(\phi\pi)+F_h(\phi\pi)],\\
F_{K^{*+}K^-}&=&\frac{2}{3}[F_a(K^*K)+F_b(K^*K)]-\frac{1}{3}[F_c(K^*K)+F_d(K^*K)], \label{KKstar1}\\
F_{K^{*0}\bar
K^0}&=&-\frac{1}{3}[F_a(K^*K)+F_b(K^*K)]-\frac{1}{3}[F_c(K^*K)+F_d(K^*K)].\label{KKstar2}
\end{eqnarray}
The form factor of $e^+e^-\to\rho\eta^{(')}$ can be written as the
combination of its $\bar nn$ and $\bar ss$ component:
\begin{eqnarray}
F_{V\eta}=\mbox{cos}\theta F_{V\eta_n}-\mbox{sin}\theta
F_{V\eta_s},\\
F_{V\eta'}=\mbox{sin}\theta F_{V\eta_n}+\mbox{cos}\theta
F_{V\eta_s},
\end{eqnarray}
where $V=\rho^0,\omega,\phi$ and
\begin{eqnarray}
F_{\rho^0\eta_n}&=&[F_a(\rho\eta_n)+F_b(\rho\eta_n)]+\frac{1}{\sqrt2}
[F_e(\rho\eta_n)+F_f(\rho\eta_n)]+\frac{5}{18}[F_g(\rho\eta_n)+F_h(\rho\eta_n)],\\
\
\
F_{\rho^0\eta_s}&=&\frac{1}{\sqrt 2}[F_e(\rho\eta_s)+F_f(\rho\eta_s)]
-\frac{\sqrt2}{6}[F_g(\rho\eta_s)+F_h(\rho\eta_s)],\\
\ \ \ \ \ \ \ \ \
F_{\omega\eta_n}&=&\frac{1}{3}[F_a(\omega\eta_n)+F_b(\omega\eta_n)]+\frac{\sqrt
2}{6}
[F_e(\omega\eta_n)+F_f(\omega\eta_n)]+\frac{5}{54}[F_g(\omega\eta_n)+F_h(\omega\eta_n)],\\
\ \ \ \ \ \ \ \ \ \
 F_{\omega\eta_s}&=&\frac{\sqrt 2}{6}
[F_e(\omega\eta_s)+F_f(\omega\eta_s)]-\frac{\sqrt2}{18}[F_g(\omega\eta_s)+F_h(\omega\eta_s)],\\
\ \ \ \ \ \ \ \ \
 F_{\phi\eta_n}&=&-\frac{1}{3}
[F_e(\phi\eta_n)+F_f(\phi\eta_n)]-\frac{5\sqrt2}{54}[F_g(\phi\eta_n)+F_h(\phi\eta_n)],\\
 \ \ \ \ \ \
F_{\phi\eta_s}&=&-\frac{2}{3}[F_a(\phi\eta_s)+F_b(\phi\eta_s)]-\frac{1}{3}
[F_e(\phi\eta_s)+F_f(\phi\eta_s)]-\frac{1}{27}[F_g(\phi\eta_s)+F_h(\phi\eta_s)].
\end{eqnarray}

\section{Numerical results and discussions}
\label{num}

\subsection{Cross section}

Making use of the  distribution amplitudes and the inputs listed
before, one can easily obtain the cross sections for the process
$e^+ e^- \to PV$. Here we would like to present the results of cross
sections at $\sqrt{s}$ =3.67 GeV and 10.58 GeV in Table~\ref{cross}
together with the data measured by CLEO-c and BaBar collaboration.
The different scenarios S1, S2, S3 denoting different transverse
momentum distribution functions, which will be discussed in the next
subsection.  As the longitudinal decay constants of the vector
mesons and the pseudoscalar meson decay constants are precisely
determined, the uncertainties from these inputs are neglected.
Therefore the uncertainties shown in Table~\ref{cross} are from the
transverse decay constants of the vector meson shown in
table~\ref{decayconstant}.

From the Eqs.~(\ref{rhopi1}) and (\ref{rhopi2}), we can see that,
if neglecting the fragmentation contribution $F_{g,h}$, the cross
sections for production of  $\rho^+\pi^-$ and $\rho^0\pi^0$ in
$e^+ e^-$ annihilation should be the same. At $\sqrt s=3.67$ GeV,
the fragmentation can not give large contribution as the
on-shellness enhancement is not strong. Thus theoretical
calculation predicts that the ratio $R_1=\frac{\sigma(e^{+} e^{-}
\to \rho^{+}\pi^{-})}{ \sigma(e^{+} e^{-} \to \rho^{0}\pi^{0})}$
should be around 1. From table~\ref{cross}, one can see that {this
prediction is consistent with the CLEO-c results.} At higher
energies, the enhancement effect becomes more important. This
effect can weaken $e^+e^-\to\rho^0\pi^0$ by about ten percents at
$\sqrt s=10.58$ GeV, relative to $e^+e^-\to\rho^+\pi^-$. If the
center mass energy is large enough, the contribution from diagrams
in Fig.~\ref{VMD} will be dominant over the other contributions.

\begin{table} \caption{Results of $e^+e^-\to VP$ cross sections at $\sqrt s=3.67$ GeV and $\sqrt s=10.58$ GeV
 using three different transverse momentum distribution functions, denoted as S1, S2 and S3 respectively.
The experimental results from CLEO-c and BaBar collaborations are
also included.}
\begin{tabular}{|c|c|c|c|c||c|c|c|c|}
\hline & \multicolumn{4}{c||}{ $\sqrt s=3.67$ GeV } & \multicolumn{4}{|c|}{ $\sqrt s=10.58$ GeV } \\
 \hline Channel & $\sigma_{S1}$(pb) &
$\sigma_{S2}$(pb)& $\sigma_{S3}$(pb) & $\sigma_{exp}$(pb)&
$\sigma_{S1}$(fb)&$\sigma_{S2}$(fb)
&$\sigma_{S3}$(fb)&$\sigma_{exp}$(fb)\\
\hline
$\rho^{+}\pi^{-}$ &$3.8^{+0.3}_{-0.2}$ &$1.9^{+0.1}_{-0.2}$ &$2.9^{+0.2}_{-0.2}$ & $4.8^{+1.5+0.5}_{-1.2-0.5}$ & $0.71^{+0.04}_{-0.04}$ &$0.55^{+0.03}_{-0.03}$& $0.68^{+0.04}_{-0.04}$&\\
$\rho^{0}\pi^{0}$ & $3.8^{+0.3}_{-0.2}$ &$1.9^{+0.1}_{-0.2}$& $2.9^{+0.2}_{-0.2}$& $3.1^{+1.0+0.4}_{-0.8-0.4}$ &$0.64^{+0.04}_{-0.04}$ &$0.50^{+0.04}_{-0.03}$& $0.62^{+0.04}_{-0.03}$&\\
$\omega \pi^{0}$ & $28.2^{+2.2}_{-2.2}$ & $13.8^{+1.1}_{-1.1}$ &$21.2^{+1.7}_{-1.6}$ & $15.2^{+2.8+1.5}_{-2.4-1.5}$ & $5.2^{+0.4}_{-0.3}$ & $4.1^{+0.5}_{-0.3}$ &$5.0^{+0.4}_{-0.3}$ &\\
$\phi \pi^{0}$& $1.2 \times 10^{-4}$& $1.2 \times 10^{-4}$& $1.2 \times 10^{-4}$ & $< 2.2$ & $2.1\times 10^{-3}$& $2.1\times 10^{-3}$& $2.1\times 10^{-3}$&\\
$K^{*+}K^{-}$& $5.6^{+0.4}_{-0.4}$ &$2.9^{+0.1}_{-0.3}$ & $4.3^{+0.3}_{-0.3}$ & $1.0^{+1.1+0.5}_{-0.7-0.5}  $ & $1.2^{+0.02}_{-0.02}$ &$0.83^{+0.05}_{-0.05}$& $1.1^{+0.0}_{-0.1}$&\\
$K^{*0}\bar{K}^{0}$ & $34.8^{+2.4}_{-2.3}$ & $17.3^{+1.2}_{-1.1}$ &$26.4^{+1.8}_{-1.8}$& $23.5^{+4.6+3.1}_{-3.9-3.1}$ &$7.1^{+0.4}_{-0.4}$  &$5.6^{+0.2}_{-0.4}$&$6.8^{+0.4}_{-0.4}$ &\\
\hline
$\rho^{0}\eta$ & $16.6^{+0.9}_{-1.0}$ & $8.1^{+0.5}_{-0.4}$ &$12.5^{+0.7}_{-0.7}$ & $10.0^{+2.2+1.0}_{-1.9-1.0}$ & $3.3^{+0.2}_{-0.2}$ & $2.4^{+0.2}_{-0.1}$& $3.1^{+0.2}_{-0.2}$&\\
$\rho^{0}\eta'$ & $8.6^{+0.6}_{-0.5}$ & $4.3^{+0.3}_{-0.3}$ &$6.6^{+0.4}_{-0.4}$& $2.1^{+4.7+0.2}_{-1.6-0.2} $ & $2.1^{+0.1}_{-0.1}$ &$1.5^{+0.1}_{-0.0}$& $2.0^{+0.1}_{-0.1}$&\\
$\omega \eta$& $1.5^{+0.1}_{-0.1}$ &$0.76^{+0.03}_{-0.07}$ &$1.1^{+0.1}_{-0.1}$& $2.3^{+1.8+0.5}_{-1.0-0.5} $ & $0.31^{+0.02}_{-0.02}$ &$0.22^{+0.02}_{-0.01}$& $0.29^{+0.02}_{-0.01}$&\\
$\omega \eta'$ & $0.79^{+0.06}_{-0.06}$ & $0.39^{+0.03}_{-0.03}$ &$0.60^{+0.04}_{-0.04}$& $<17.1$ & $0.20^{+0.01}_{-0.02}$ &$0.14^{+0.01}_{-0.01}$& $0.18^{+0.012}_{-0.01}$&\\
$\phi \eta $& $19.1^{+1.1}_{-1.1}$ & $9.6^{+0.6}_{-0.6}$ & $14.6^{+0.8}_{-0.9}$& $2.1^{+1.9+0.2}_{-1.2-0.2}  $ & $4.3^{+0.2}_{-0.2}$ &$3.3^{+0.2}_{-0.2}$& $4.1^{+0.1}_{-0.2}$&$2.9 ^{+0.5+0.1}_{-0.5-0.1}$\\
$\phi \eta'$ & $22.6^{+1.4}_{-1.3}$ & $11.5 ^{+0.8}_{-0.7}$ &$17.4 ^{+1.1}_{-1.0}$&$< 12.6$ & $5.8 ^{+0.3}_{-0.3}$ &$4.4 ^{+0.2}_{-0.3}$&$5.4 ^{+0.4}_{-0.3}$ &\\
\hline
\end{tabular}  \label{cross}
\end {table}

The process $e^+e^- \to K^* K$ has previously been calculated in
PQCD ($k_T$ factorization) and has been shown to give correct
order of magnitude for the form factors \cite{CKM05}. But they
assume SU(3) symmetry using asymptotic wave functions.
 In order to show the $SU(3)$ symmetry breaking effect in
$e^+e^-\to K^*K$, we define the ratio:
\begin{eqnarray}
R_2=\frac{\sigma(e^+e^-\to K^{*0}\bar K^0)}{\sigma(e^+e^-\to
K^{*+}K^-)}=\bigg|\frac{1+\frac{F_c+F_d}{F_a+F_b}}{2-\frac{F_c+F_d}{F_a+F_b}}\bigg|^2.
\end{eqnarray}
If we assume that $SU(3)$ symmetry works well, then the light cone
distribution amplitude of $K$ and $K^*$ is completely symmetric
under the exchange of the momentum fractions of quark and
anti-quark. We will have $F_a+F_b=F_c+F_d$, then $R_2=4$ can be
derived directly from Eq.~(\ref{KKstar1}) and (\ref{KKstar2}). One
of the SU(3) symmetry breaking effects is that the $s$ quark is
heavier than $n(=u,d)$ quark and carries more momentum in the final
state light $K^{(*)}$ meson. The gluon which generates $\bar ss$ is
harder than the $\bar nn$ generator, then the former coupling
constant is smaller due to the more off-shell gluon. Consequently
this leads to a smaller contribution to the form factor $|F_a+F_b|$
than $|F_c+F_d|$. Therefore $R_2$ is larger than 4. Using the cross
sections listed in Table~\ref{cross},  we obtain our result for
$R_2$:
\begin{eqnarray} R_2=6.0,
\end{eqnarray} where only the central value is given.
The CLEO-c results indicate that there is a large deviation from the
$SU(3)$ limit \cite{ExpVP3}:
\begin{eqnarray}
R_2=23.5^{+17.1}_{-26.1}\pm12.2.
\end{eqnarray}
The central value of the experimental results for $R_2$ seems too
large, but as the uncertainties are also large, our result could be
consistent with results from CLEO-c collaboration.

For the processes involving $\eta^{(')}$ such as the process
$e^+e^-\to\phi\eta$, we find that the gluonic contribution is
around one percent to the total cross section at $\sqrt s=10.58$
GeV. This conclusion is consistent with the study on the $B\to
\eta^{(')}$ form factor \cite{GluonicCKL}. The cross sections of
$e^{+} e^{-} \to \rho(\omega) \eta^{(\prime)}$ and $\omega\pi^0$
at $\sqrt s=3.67$ GeV calculated in $k_{T}$ factorization are
consistent with the experimental values. The result for $e^+e^-\to
\phi\eta$ at $\sqrt s=10.58$ GeV is also consistent with
experimental data. This indicates that $k_{T}$ factorization is an
effective method to deal with the infrared divergences in
exclusive processes.

The measurement of cross sections at different center  mass energy
$\sqrt{s}$ can shed light on the $s$ dependence. This dependence is
expected as $1/s^3$ \cite{sdependence3} or $1/s^4$\cite{pQCDBL,
sdependence4}. Our results displayed in Table~\ref{cross} at two
different scales $\sqrt{s}=10.58$ GeV and 3.67 GeV seem to favor the
$1/s^3$ scaling. It should be noticed that we neglected the
$Q=\sqrt{s}$ dependence of the light cone wave function and
next-to-leading order contributions in our calculation. Therefore
the $s$ dependence study of the cross sections is not a complete
one.

As the quark or the gluon could be on-shell, we expect the
amplitudes receive an imaginary part which is similar with the
exclusive $B$ decays~\cite{PQCDBdecay,Bdecayversus}. The imaginary
part in $e^+e^-\to \phi\eta$ at $\sqrt s=10.58$ GeV is about twice
as large as the real part in magnitude and a large strong phase is
consequently generated.  The contributions from Fig.~\ref{gluonic}
are small; Contributions from Fig.~\ref{VMD} are small and real; The
four diagrams in Fig.~\ref{quark} give comparable contributions
which are the main origin of the imaginary part. Unlike the B
decays, strong phase here does not make any physical meaning, since
there is no electroweak phase for interference.

From Table~\ref{cross}, we can see that at $\sqrt s=10.58$ GeV,
the cross sections for many processes, especially $e^+e^-\to
K^{0*}\bar K^0$ and $e^+e^-\to\phi \eta'$, are large enough to be
detected. We suggest the experimentalists to measure these
channels.

In the above, we only concentrate on the exclusive production of a
vector and a pseudoscalar. Applications to $PP$ and $VV$
productions are straightforward. The diagrams in Fig.~\ref{quark}
will give dominant contributions, where the final state must have
negative charge conjugation quantum number $C=-1$. Then only three
channels for $PP$ is allowed through one photon annihilation:
$e^+e^-\to\pi^+\pi^-$, $e^+e^-\to K^+K^-$ and $e^+e^-\to K^0\bar
K^0$. If $U$-spin is well respected, $d$ and $s$ quarks are
symmetric in $K^0$ and the cross section of $e^+e^-\to K^0\bar
K^0$ is zero. The non-zero result for $e^+e^-\to K^0\bar K^0$ can
reflect the size of $U$-spin symmetry breaking. For production of
$VV$, the analysis is similar. Two flavor singlet vector mesons
can not be produced through one photon annihilation diagrams
either, but these productions could receive large additional
contributions \cite{fragementation}.

\subsection{Theoretical Uncertainties}

One of the major uncertainties in our computations comes from the
distribution amplitudes for the pseudoscalar and vector mesons. The
dependence on the longitudinal distribution amplitudes has been
studied intensively in the exclusive B decays \cite{Kurimuto}. They
will give 10-20\% uncertainties here too. In the following, we will
focus on the transverse momentum distribution. In PQCD approach, the
intrinsic transverse momentum is taken into account. The resummation
of large double logarithms results in the Sudakov factor which
suppresses the large $b$ region's contribution. As we can see from
Eq.~(\ref{transverse}), the transverse momentum distribution
function also suppress the contribution from the large $b$ region.
For momentum transfer of a few GeV, the transverse momentum
distribution function damps more than the Sudakov factor
\cite{transverseDA}. This suppression makes PQCD approach more
self-consistent. So we expect that there is an obvious suppression
for the production rate of $e^+e^-\to VP$ at $\sqrt s=3.67$ GeV if
the transverse momentum distribution amplitude is taken into
account. However, at present, it is still lack of first-principle
study on the intrinsic transverse momentum distribution.   The
simple form is chosen as the Gaussian form discussed in
Eq.~(\ref{transverse}) or the following one,
\begin{eqnarray}
\Sigma(x,b)=\mbox{exp}\left[-\frac{x(1-x)b^2}{4a^2}\right],\label{transverse2}
\end{eqnarray}
where $a$ is the transverse size parameter as $\beta$.  As a simple
test, we can choose $a=1$ which is consistent with the value used in
\cite{Kurimuto}. Comparing with the form in Eq.~(\ref{transverse}),
we can see that at $x=1/2$, the two different forms coincides. But
for small or large $x$, the second form can not give the same strong
suppression as the first form (Eq.~(\ref{transverse})). We can
expect the suppression of the results of taking the second form is
less effective than the first one. In Table~\ref{cross}, we give
three different kinds of results: without the intrinsic momentum
distribution (denoted as S1), i.e. $\Sigma=1$; with the first
distribution as Eq.~(\ref{transverse}) (denoted as S2); with the
second kind as Eq.~(\ref{transverse2}) (denoted as S3). Comparing
the different results in Table~\ref{cross}, we find that at small
center mass energy $\sqrt s=3.67$ GeV the suppression from
transverse momentum distribution is more effective: the suppression
is $50\%$ for S2 and $20\%$ for S3. Since the results depend on the
explicit form of transverse momentum distribution, more experimental
results are needed.

In this calculation, we only present the leading order
calculations. The complete next-to-leading order calculations are
much more complicated~\cite{NLO}. For a simple estimate of the
size of the next-to-leading order contribution, we use the
traditional method varying $\Lambda_{QCD}$ and factorization scale
$t$ in Eq.~(\ref{hardscale1}), (\ref{hardscale2}) and
(\ref{hardscale3}): $\Lambda_{QCD}=(0.25\pm0.05)$ GeV; changing
hard scale $t$ from $0.75t$ to $1.25t$ (not changing $1/b_i$). We
find that our results are not sensitive to these changes. This
implies that the next-to-leading order contribution is probably
not very large.

\section{Conclusions}

In this paper, we have studied the exclusive processes $e^+e^-\to
VP$ in PQCD approach based on the $k_T$ factorization. We give three
different kinds of results corresponding to different  transverse
momentum distribution functions.  With the proper distribution
function, our results can be consistent with the experimental
results. The two different transverse momentum distribution
functions S2 and S3 can give about $50 \% $ and $20 \% $ suppression
respectively at center mass energy $\sqrt{s}=3.67 \rm{{GeV}}$. We
have included the gluonic contribution for the processes involving
$\eta^{(')}$ meson whose effect is found tiny. We have also included
the contribution in which the flavor singlet vector meson is
produced by an additional photon. This contribution could be
neglected at center mass energy $\sqrt{s}=3.67 \rm{{GeV}}$, while
these diagrams could induce about $10\%$ difference between
$e^+e^-\to\rho^0\pi^0$ and $e^+e^-\to\rho^+\pi^-$ at $\sqrt s=10.58$
GeV. The $s$ dependence of the cross section has been directly
studied which indicates that the $1/s^3$ scaling is more favored
than $1/s^4$.

\section*{Acknowledgement}
We would like to thank Y. Jia and Y.L. Shen for useful discussions
and suggestions. This work is partly supported by National Science
Foundation of China under Grant No.~10475085 and 10625525.

\end{document}